\documentclass[aps,twocolumn]{revtex4}

\usepackage{mathrsfs}
\usepackage{amsmath,amssymb,amsfonts}

\newcommand{\be}{\begin{equation}}
\newcommand{\ee}{\end{equation}}
\newcommand{\bra}{\left\langle}
\newcommand{\ket}{\right\rangle}

\begin{document}

\title{The chiral condensate in holographic models of QCD}

\author{Aleksey Cherman}
\email{alekseyc@physics.umd.edu}

\author{Thomas D. Cohen}
\email{cohen@physics.umd.edu}

\author{Elizabeth S. Werbos}
\email{ewerbos@physics.umd.edu}

\affiliation{Department of Physics, University of Maryland,
College Park, MD 20742-4111}

\begin{abstract}
Bottom-up holographic models of QCD, inspired by the AdS/CFT
correspondence, have shown a remarkable degree of phenomenological
success.  However, they rely on a number of bold assumptions.  We
investigate the reliability of one of the key assumptions, which
involves matching the parameters of these models to QCD at high 4D
momentum $q^2$ and renormalization scale $\mu^2$. We show that this
leads to phenomenological and theoretical inconsistencies for
scale-dependent quantities such as $\bra \bar{q}q\ket$.
\end{abstract}

\maketitle

There are still no systematic analytic tools to study the
strong-coupling dynamics of QCD, except for models that probe certain
limited classes of observables.  Thus, for example, one can use chiral
perturbation theory for some low-energy observables, but for more
general ones one is forced to resort to more phenomenological
approaches such as chiral soliton models.  In recent years,
`bottom-up' holographic models of QCD have emerged as another approach
to the low energy phenomenology of QCD, and have attracted
considerable
interest\cite{EKSS,DRP1,DRP2,Ghoroku,BoschiFilho:2002vd,BoschiFilho:2002ta,KatzTensorMesons,HirnRuizSanz,ErlichKribsLow,KKSS,BrodskydeTeramond,ShockWu,EvansTedderWaterson,CsakiReece,ShockBackreacted,Schafer,Forkel,KweeLebed,ColangeloEtAl}.

In these models, QCD in the large $N_c$ limit is taken to be dual to a
classical 5D theory in a curved space, and the parameters of the 5D
model are matched to their corresponding values in large $N_c$ QCD\cite{tHooft, WittenNc},
with the field content of the 5D models chosen to match the low energy
chiral symmetry of QCD.  In contrast to approaches like chiral
perturbation theory, these models allow the computation of meson
spectra and couplings, at least in principle.  Even very simple 5D
models seem to show a remarkable phenomenological success when
compared to data (see, for instance, ref.~\cite{EKSS,DRP1}).

There are, however, some phenomenological puzzles with the models.
For example, quark masses as extracted by ref.~\cite{EKSS} and
ref.~\cite{DRP2} are different by a factor of four. This
inconsistency, which we address in this Letter, is in fact symptomatic
of issues which are more serious than they may first appear. The
resolution of these issues, associated with the treatment of operators
which are scale dependent in QCD in holographic models, will shed
light on the the matching of the 5D models to QCD.

Holographic models of QCD are motivated by the conjectured dualities
between some types of gauge theories and higher-dimensional gravity
theories.  The most well-known such duality is the AdS/CFT
correspondence \cite{Maldacena,GubserEtAl,Witten}, where a conformal
field theory, ${\cal N} = 4$ $SU(N_c)$ super Yang-Mills theory in the
large $N_c$ limit, is dual to a type IIB string theory on
$AdS_5\times\mathbf{S}^5$, where $AdS_5$ is 5D Anti-DeSitter space, and $\mathbf{S}^5$ is the 5-sphere.  When the 't Hooft coupling of the field
theory is large, the $AdS_5\times\mathbf{S}^5$ physics is described by weakly-curved classical supergravity.  The CFT lives on the boundary of $AdS_5$,
in the sense that each operator $\mathcal{O}$ of the CFT is identified
with a bulk field $\phi^{\mathcal{O}}$ in $AdS_5\times\mathbf{S}^5$,
the boundary values of the bulk fields $\phi^{\mathcal{O}}_{0}$ are
the sources for the CFT operators, and the supergravity partition
function $Z_{\mathrm{SG}}$ is identified as the generating functional
of the CFT correlation functions:
\begin{equation}
\label{ActionMatching}
Z_{\textrm{SG}}[\phi^{\mathcal{O}}_{0}]=\int_{\phi^{\mathcal{O}} \rightarrow \phi^{\mathcal{O}}_{0}}{D\phi^{\mathcal{O}}\: e^{-S_{SG}[\phi^{\mathcal{O}}] }}
=\left< e^{-\int_{\partial \mathrm{AdS}}{\phi^{\mathcal{O}}_{0} \mathcal{O}}} \right>_{\mathrm{CFT}} .
\end{equation}
CFT correlation functions in the strongly-coupled domain can be
calculated by evaluating the supergravity action $S_{SG}$ on the
classical solution for fields $\phi^{\mathcal{O}}$ that approaches a
specified boundary value $\phi^{\mathcal{O}}_0$, and
taking functional derivatives with respect to
$\phi^{\mathcal{O}}_0$\cite{GubserEtAl,Witten}.

The bottom-up approach to holographic QCD generally consists of
guessing a 5D background and field content that captures some aspects
of large $N_c$ QCD for some observables of interest.  In AdS/CFT, the
conformal invariance of the CFT corresponds to the fact that
coordinate rescaling is an isometry of AdS.  QCD is approximately
conformal at high energies (due to asymptotic freedom), but not at low
energies (due to confinement).  This means that an unmodified $AdS_5$
background can not capture the essential features of QCD, and the
$AdS_5$ space must be modified in some way.  Generally, the
holographic QCD models on the market use an asymptotically $AdS_5$
background to reflect the fact that QCD is approximately conformal at
high energies.  The deep bulk region is then modified (that is, cut
off in some way) in order to model confinement.  For example, in hard
wall models\cite{EKSS,DRP1,DRP2,KatzTensorMesons,HirnRuizSanz,ErlichKribsLow,ShockWu,Forkel}, confinement is
simulated by cutting off the AdS space at some finite radius by hand.  In soft-wall
models, a dilaton field is turned in the bulk and tuned to smoothly cut off the AdS space to
produce a linear Regge meson mass spectrum \cite{KKSS,Forkel,KweeLebed,ColangeloEtAl}, as opposed to
the quadratic spectrum that is generally seen in hard-wall models.
Other models take into account the back-reaction of the bulk fields on
the metric, which can dynamically cut off the AdS space
\cite{CsakiReece,ShockBackreacted}.

A well-known general theoretical issue with all such models is that in
the regime where QCD is approximately conformal, it is weakly coupled
due to asymptotic freedom.  As a result, it is far from obvious that
the use of a classical, weakly curved 5D background is justified,
since one generally expects that a systematic holographic dual of QCD
should be a string theory on some highly curved space (as noted for
instance in ref.~\cite{KKSS}).  However, despite this possible
problem, it is interesting to try to investigate whether a classical
5D background might serve as a phenomenologically useful ad hoc
approximation to a holographic dual of QCD, and many bottom-up models
appear to show a remarkable agreement with experimental data.

The AdS/CFT dictionary\cite{GubserEtAl,Witten} is used in bottom-up
models to dictate the bulk field content: a p-form 4D QFT operator
$\mathcal{O}$ with scaling dimension $\Delta$ corresponds to a p-form
bulk field with mass $m_{5}^2=(\Delta -p)(\Delta+p-4)$.  Conserved
currents in the QFT correspond to gauge fields in the bulk, with
$m_{5}^2=0$.  In bottom-up holographic models, it is assumed that this
dictionary remains valid even when one modifies the 5D background away
from $AdS_5$.  It is not obvious that this should be justified,
because for instance while the scaling dimension of operators is
well-defined in a CFT, in QCD the scaling dimensions of most operators
receive scale-dependent corrections proportional to $\alpha_s$. That
is, most operators in QCD have scale-dependent anomalous dimensions.
An exception is operators that correspond to conserved currents, i.e.,
those associated with vector and axial currents, and the stress-energy
tensor, whose anomalous dimensions are zero.  However, in an exploratory spirit holographic models of QCD assume that
the dictionary remains valid.

In order to model some observables of interest, one must choose a set of
4D operators and corresponding 5D fields that probe the relevant
physics.  Of course, QCD has an infinite number of operators with the
same quantum numbers, and due to operator mixing and the lack of any obvious suppression scale, an arbitrarily large
subset of them can contribute to any given process\cite{Glozman}.  As
an ad hoc approximation, to obtain tractable holographic models one simply
chooses a minimal set of lowest-dimension operators to probe the observables one is
interested in.  The parameters of the resulting 5D
models are matched using the AdS/CFT dictionary to large $N_c$ QCD in the UV regime,
where the asymptotic freedom of QCD allows reliable perturbative
calculations.

Given the many bold assumptions that are necessary to construct
holographic models of QCD, their phenomenological success is
remarkable, and may suggest that the assumptions are more reliable
than might be expected.  It is natural to wonder if there is anything
in these models that can test the reliability of the assumptions.  In
this Letter, we show that the ad hoc treatment of scale-dependent
operators in holographic models of QCD can lead to serious
phenomenological and theoretical problems.  For definiteness, we work
in the simple hard-wall model of ref.~\cite{EKSS} for chiral symmetry
breaking, and focus on the behavior of the quark condensate $\bra
\bar{q}q\ket$ and its source $m_q$.  We show that to match the large
$N_c$ scaling of QCD quantities, the operator-field mapping for the
quark condensate needs to be properly normalized, and this
normalization affects the value of the quark mass and the quark
condensate.  Next, we argue that the values of the quark mass and the chiral condensate obtained
in the model from matching to data are inconsistent with the matching
of the 5D models to QCD in the asymptotically UV region, and discuss
the implications of the analysis for the treatment of scale dependent
operators in bottom-up holographic models of QCD.

\section{The model}
Our analysis is in the simple model of ref. \cite{EKSS}, which we
describe in this section.  QCD has an $SU(N_f)_L\times SU(N_f)_R$ flavor symmetry in the chiral limit, and the model focuses on the lowest-dimensional
operators important to chiral dynamics. These are the left and
right handed quark currents $\bar{q}_{L}\gamma_{\mu} t^{a} q_{L}$ and
$\bar{q}_{R}\gamma_{\mu} t^{a} q_{R}$, where $t^{a}$ are the generators of $SU(N_f)$.  They are associated with the 5D gauge
fields $A^{a}_{L}$ and $A^{a}_{R}$ using the AdS/CFT dictionary.  We will work in the $N_f=2$ limit in this Letter.

To model chiral symmetry breaking, we include the operator
$\bar{q}_R q_L$, which acquires a vacuum expectation value, and associate it with a massive bulk scalar field
$\phi$. (The dictionary between $\bar{q}_R q_L$ and $\phi$ includes a
factor of $1/z$ due to the dimension of $\bar{q}_R q_L$.)  The field
content of the model and the dictionary are given in Table 1.  Note
that in assigning these 5D masses it is assumed that the scaling
dimensions of the operators are purely classical.  Note that this is
not the case for $\bar{q}_R q_L$ in QCD, so that this is an ad hoc approximation.  
\begin{table}
\label{tab:fields}
\begin{tabular}{ccccc}
\hline 4D Operators & 5D Fields & $p$ & $\Delta$ &
$(m_5)^2$\\ \hline $\bar{q}_L\gamma^\mu t^aq_L$ & $A_{L
\mu}^a$&1&3&0\\ $\bar{q}_R\gamma^\mu t^aq_R$ & $A_{R \mu}^a$&1&3&0\\
$\bar{q}_R^a q_L^b$ & $\frac{2}{z}\phi^{a b}$&0&3&-3\\
\hline
\end{tabular}
\caption{Field content and dictionary of the model.}
\end{table}
The model uses an AdS space with a hard-wall cutoff in the IR as the
holographic 5D background.  The metric is
\begin{equation}
ds^2=g_{\mu \nu} dx^{\mu}dx^{\nu}=\frac{L^2}{z^2} (-dz^2 + dx^{i} dx_{i}),
\end{equation}
where L is the radius of the AdS space, which will be set to unity in
the rest of this Letter, and $z \in [0,z_m]$.  Holographic
calculations generally use $z=\epsilon$ as a UV
regulator\cite{Skenderis}, with $\epsilon \rightarrow 0$ at the end of
the computations. We use the convention that Latin indices take values
$0,1,2,3$, and Greek indices take values $z,0,1,2,3$.  This choice of
metric corresponds to the assumption that QCD, which lives on the
$z=0$ boundary, remains conformal until confinement suddenly sets in
at an energy scale of order $1/z_m$.  While this is clearly a drastic
oversimplification, it does not affect our general conclusions.

In terms of the bulk fields that we have defined above, the classical
bulk action can be written as
\begin{equation}
S = \int d^5x \sqrt{g}\mathrm{Tr}\left\{|D\phi|^2 + 3|\phi|^2 - \frac{1}{4g_5^2}\left(F_L^2+F_R^2\right)\right\},
\end{equation}
with $D_\mu \phi = \partial_\mu \phi - iA_{L\mu}X+iXA_{R\mu}$,
$A_{L,R}=A_{L,R}^a t^a$, and $F_{\mu\nu} = \partial_\mu
A_\nu-\partial_\nu A_\mu - i \left[A_\mu,A_\nu\right]$.  The gauge
coupling $g_5$ can be determined from matching the two-point vector
correlation function to the leading term in the QCD operator product
expansion for the vector two-point function.  To do this, one can
evaluate the 5D action on solutions of the equations of motion of the
vector field ($V=A_L+A_R$), assuming that the 5D side can be treated
classically, and take two derivatives with respect to the boundary
sources to obtain the vector two-point function.  The vector two-point
correlation function is $\int{d^{4}x e^{ipx} \bra J_{\mu}(x)
J_{\nu}(0)\ket}=(q_{\mu}q_{\nu} -q^2 g_{\mu \nu})\Pi_{V} (q^2)$, and
it can be shown that near the boundary, which corresponds to large
scales in the boundary field theory, one obtains\cite{EKSS} \be \Pi_V
(q^2)=-\frac{1}{2g_5^2}\log{(q^2)}\;, \ee where $q$ is the 4D
momentum.  This can be matched to the corresponding QCD expression,
coming from a quark bubble and valid at high $q^2$, where QCD is
weakly coupled due to asymptotic freedom: \be \Pi_V
(q^2)=-\frac{N_c}{24\pi^2}\log{(q^2)}\;.  \ee Matching the holographic
model to QCD implies that $g_5^2=12\pi^2/N_c$.

To incorporate chiral symmetry breaking, it is necessary to connect
the scalar field $\phi$, which is dual to $\bar{q}_R q_L$, to the
chiral condensate $\bra \bar{q}_R q_L \ket = \Sigma=\sigma \mathbf{1}$ and the quark mass
$M=m_q\mathbf{1}$, which is the source of $\bar{q}_R q_L$ in QCD.
There is a straightforward prescription for this in AdS/CFT.  As shown
by Klebanov and Witten\cite{KlebanovWitten}, the near-boundary
solution of the equations of motion for a field $\phi^{\mathcal{O}}$
dual to an operator $\mathcal{O}$ with dimension $\Delta$ is given by
\be
 \phi^{\mathcal{O}}(z,q)= z^{4-\Delta}\phi^{\mathcal{O}}_{0}(q) +
z^{\Delta} \frac{\bra \mathcal{O}\ket} {2\Delta - 4} \;, 
\ee 
where
$\phi^{\mathcal{O}}_{0}(q)$ is the source for $\mathcal{O}$, and $\bra
\mathcal{O}\ket =\delta \log{Z}/\delta \phi^{\mathcal{O}}_{0}(q) $ is
the one-point correlation function.

Before applying this to our problem, we must address a slightly subtle
normalization issue in the association of operators with fields in
holographic models.  In a field theory with an operator $\mathcal{O}$
and a source $J$, one always has the trivial freedom to redefine
$\mathcal{O} \rightarrow a \mathcal{O}$ and $J \rightarrow J/a$, so
that $J \mathcal{O}$ is unchanged.  In holographic models defined by
actions such as Eq.~\ref{ActionMatching}, this amounts to the freedom
to take $\mathcal{O} \rightarrow a \mathcal{O}$ and
$\phi^{\mathcal{O}}_{0} \rightarrow
\phi^{\mathcal{O}}_{0}/a$\cite{SonStephanov}.  Many treatments of
holographic models of QCD implicitly take $a=1$.  However, as we are
about to show, this is not generally correct, and the issue of
operator/source normalization turns out to have significant
implications for holographic models such as the one in this Letter.

We specialize the discussion above to our case with
$\mathcal{O}=\bar{q}_R q_L$ and $\Delta=3$, keeping the normalization
parameter $a$ explicit in the equations.  Using the dictionary $\lim_{z\rightarrow 0} \frac{2}{z} \phi(z,q) = a M$, we can
write the $q=0$ classical solution for $\phi$ as \be
\label{ZeroQ}
\phi(z)= \frac{1}{2} a M z + \frac{\Sigma}{2 a} z^3,
\ee
where $M=m_q \mathbf{1}$ and $\Sigma$ is $\Sigma=\bra \bar{q}_L q_R \ket = \sigma \mathbf{1}$.

It is now clear how the seemingly trivial issue of source and operator
normalization becomes important.  If we take $a=1$, as was done in
ref.~ \cite{EKSS} implicitly, then Eq.~\ref{ZeroQ} has incommensurate
$N_c$ scaling: the first term scales as $N_c^0$, while the second
scales as $N_c^1$.  It is also not hard to show that with $a=1$ in
this model, the $\rho/a_1$ mass splitting scales $N_c^{1/6}$, in
direct contradiction with large $N_c$ QCD.  To be consistent with
large $N_c$ QCD, we must take $a \sim N_c^{1/2}$ rather than $a=1$, so that the $\rho/a_1$ mass splitting scales as $N_c^{0}$\cite{WittenNc} .  Of course, this should not be surprising: in large $N_c$ QCD, the normalization of operators explicitly depends on $N_c$, so the same must be true in the holographic approach. 

To complete the description of the holographic model, one must specify
the values of $\sigma,m_q$ and $a$.  The chiral condensate $\sigma$
comes from IR physics, and in this model it can be taken to be an
input parameter.  In fact, the structure of the model fixes $\sigma/a$ and $a m_q$.  To determine $\sigma/a$, one can use the relation\cite{EKSS}
\be
f_{\pi}^2=\left. -\frac{1}{g_5^2}  \frac{\partial A(z)}{\partial z}\right|_{z=\epsilon}\;,
\ee
where A(z) is the transverse part of the axial vector current ($A=A_L+A_R$) at $q=0$, which satisfies the equation
\be
\partial_z \left(z^{-1} \partial_z A(z) \right) - \frac{g_5^2 \sigma^2 z^3}{a^2} A(z)=0\;,
\ee
which is shown in the chiral limit $m_q=0$ for simplicity.  Since this model obeys the Gell-Mann-Oakes-Renner (GOR)
relation\cite{EKSS} $m_{\pi}^2 f_{\pi}^2 = 2 m_q \sigma$, it is possible to calculate $a m_q$ and $\sigma/a$ by 
fitting $m_{\pi}^2$ and $f_{\pi}^2$ to data.

To complete the specification of the model, it is necessary to determine the normalization parameter $a$. We will compute
$a$ by matching $a$ to QCD explicitly, as was first done in
ref.~\cite{DRP2}, and show that this leads to the correct $N_c$
scaling. To do this, we will compare the two-point functions for
$\bar{q}_R q_L$ in the holographic models and in the asymptotic regime
in QCD, using an identical procedure to the one that was followed for
$g_5^2$ in the vector sector.  We will see that this brings up a troubling issue involving the matching of the holographic model to QCD.

\section{Matching of the scalar section to QCD}
The equation of motion of the scalar field for general 4D momentum $q$ is
\be
\label{EoM}
\partial_z (z^{-3} \partial_z \phi(z,q)) + \frac{3+z^2 q^2}{z^5}
\phi(z,q) = 0, \ee The solution involves Bessel functions, and can be
matched for small $q$ to Eq.~\ref{ZeroQ}.  To compute the two point
correlation function, we evaluate the scalar field action on a
solution $\phi_{\mathrm{cl}}$ of Eq.~\ref{EoM}, which leaves a
boundary term:
\be S[\phi_{\mathrm{cl}}]=\int{d^{4}x \left(z^{-3}
\phi_{\mathrm{cl}}(z,q)\partial_z \phi_{\mathrm{cl}}(z,q)
\right)_{z={\epsilon}}}\;.
\ee
To find the two point correlation
function, we can now take two functional derivatives with respect to
the source $\lim_{z\rightarrow 0} \frac{2}{z} \phi(z,q) = a M$, and in
the large $q$ limit we find that
\be
\label{AdSa}
 \int{d^{4}x e^{ipx} \bra
\bar{q}q(x) \bar{q}q(0)\ket} = \frac{a^2}{2}q^2 \log{(q^2 L^2 )} + \ldots \;,
\ee
where we have suppressed contact terms.  This can be compared to the
QCD result for large euclidean momentum $q^2$ and renormalization
scale $\mu$:
\be
\label{QCDa}
 \int{d^{4}x e^{ipx} \bra \bar{q}q(x) \bar{q}q(0)\ket}
= \frac{N_c}{8 \pi^2}q^2 \log{(q^2/\mu^2)}+ \ldots \;.
\ee
This implies that
$a=\sqrt{N_c}/2\pi$, which matches the $N_c$ scaling we expected on
general grounds.   This identification does not depend on any matching between $\mu$ and $L$, since both of them can be reabsorbed into the subleading terms in the equations above.  Following the fitting procedure of ref.~\cite{EKSS}, but using $a$ as extracted above yields $m_q=8.3\:
\mathrm{MeV}$ and $\sigma=(213\:\mathrm{MeV})^3$, which differs from
the results one obtains with $a=1$, where $m_q=2.29\: \mathrm{MeV}$
and $\sigma=(327\:\mathrm{MeV})^3$\cite{EKSS}.

Of course, even though $m_q$ and $\sigma$ are determined by matching
to experimental data, they have a very specific interpretation in the
holographic model that can be checked against QCD: $M=m_q \mathbf{1}$
is the source of the QCD operator $\bra \bar{q}_R q_{L} \ket$.

This presents a serious theoretical problem, since in QCD the value of $m_q$ and $\sigma$
are scale dependent.  However, the quantities $a m_q$ and $\sigma /a$ in the holographic model are fixed by matching fitting to the GOR relation and relating $f_{\pi}^2$ and $\sigma$.  The computation of $a$ above then determines $m_q$ and $\sigma$, and since $a$ does not depend on a renormalization scale, neither do $m_q$ and $\sigma$, in conflict with the identification of $m_q$ as the source of the QCD operator $\bar{q}_R q_L$.

This situation is actually rather common in phenomenological models of QCD, where one
probes various observables that are scale-dependent in QCD in models
where they do not depend on the scale in any systematic way.  Examples
of this include the treatment of structure functions in bag models, and the
treatment of chiral condensates in Nambu-Jona-Lasinio (NJL) models\cite{NJL}.
The phenomenological models are generally taken to be at some
``natural scale,'' generally about $1\: GeV$, and the observables
computed in the models are taken to correspond to QCD quantities
evaluated at that scale.  This is the interpretation taken in
ref.~\cite{DRP2} for $m_q$ and $\sigma$.

However, while this phenomenological approach may be reasonable in
models such as NJL model, it is not consistent with the
general structure of holographic models of QCD.  One of the greatest
attractions of holographic models of QCD is that they can be matched
to QCD.  The features and parameters of the holographic models are
generally taken from the AdS/CFT dictionary, so that the 5D model is
matched to QCD on the AdS boundary.  To match the parameters of the
holographic models to QCD, it is necessary to compute correlation
functions in the bulk, and then match them on the AdS boundary at high
4D momentum $q^2$ to the equivalent QCD parameters evaluated at the
same high $q^2$ and renormalization scale $\mu^2$.  However, in the
matching procedure for $a$ above, which is used to find $m_q$, the renormalization scale $\mu$ does not
appear explicitly on the 5D side, since the 5D model is classical, so that the scale to which the holographic computations are supposed to correspond to is not obvious.  

Of course, in holographic models it is generally assumed that $1/z$ plays the role of $\mu$.  However, this
does not resolve the issues with scale dependent quantities
like $m_q$ and $\sigma$.  The issue is due to the fact that the AdS/CFT dictionary
relates fields on the AdS $z=0$ boundary to QCD quantities in the UV.
The matching to QCD is done at asymptotically high scales,
where it is weakly coupled, and $\mu \rightarrow \infty$, in
accordance with the identification $1/z \sim \mu$ on the holographic side.  However, in QCD\cite{PeskinSchroeder}, as $\mu \rightarrow \infty$ the quark mass $m_q$ runs to zero and chiral condensate $\sigma$ runs to infinity.  Consistency with QCD then implies
that in the holographic model $m_q$, which is fixed on the $z=0$ boundary,
should \emph{also} be zero, while $\sigma$ in the model must diverge because of the GOR relation.  
 
This is clearly inconsistent with the phenomenology of the model, which \emph{requires} that $\sigma \neq \infty$ in order to have a finite splitting between the $\rho$ and the $a_1$ mesons. We note, moreover, that this problem does not go away in the chiral limit of $m_{\pi}^2=0$, since $\sigma$ still diverges.

\section{Discussion}
Our analysis above should not be surprising: the construction of the
holographic model required a number of ad hoc assumptions that are
clearly connected to this issue. The behavior of scale-dependent
quantities like the chiral condensate is an explicit probe of the
self-consistency of the assumptions.  Clearly, if one wants to match the key features of large $N_c$ QCD in a consistent way, it is essential to capture the scale dependence of QCD on the 5D side of the model.  This amounts to trying
to improve on the the ad hoc approximations involved in the
construction of the 5D model.

It is well known that the $\bra \bar{q}_R q_L\ket$ has a
scale-dependent anomalous dimension $\delta(\mu)$ proportional to
$\alpha_s$ to leading order.  Presumably, incorporating such an an
anomalous dimension would change the 5D mass of the scalar field.
Since the anomalous dimension depends on the scale $\mu \sim 1/z$, it
is reasonable to make the 5D mass in a holographic model depend on
$z$, $m_5^2 = m_{5}^2(z)$, with the constraint that
$\lim_{z\rightarrow 0} m_{5}^{2}(z)= - 3$.  Matching the 5D model to QCD
for $m_q$ would then amount to demanding that $m_q$ in the holographic
model obey the same renormalization group equation as the quark mass
in QCD\cite{Shuryak}. The anomalous dimension is proportional to the running
coupling $\alpha_s$, suggesting that one must also modify the 5D
background to allow the $\alpha_s$ to run\cite{CsakiReece}.  Also, one can include a more general potential in the action for the field $\phi$\cite{Gubser}, at the price of increasing the number of parameters in the model.

Although we have focused our analysis on the simple model of
ref.~\cite{EKSS}, the problems with scale dependence apply rather
broadly to bottom-up holographic models of QCD, which as yet have not
treated scale-dependence of quantities like $m_q$ and $\sigma$
consistently.  It is an open question as to whether
position-dependent 5D masses and more realistic 5D geometries could
make the treatment of the chiral condensate in holographic models
consistent with QCD.  

The analysis here explicitly demonstrates that potentially serious
consequences arise for scale-dependent quantities when one attempts to match holographic models to
QCD in the weakly coupled asymptotic region.  One can try to avoid the issues of matching to QCD in its weakly coupled regime by imposing a UV cutoff on the holographic models, as for instance in refs.~\cite{EvansShock, EvansTedder}.  However, refs.~\cite{EvansShock, EvansTedder} showed this that a fit to data with the UV cutoff as an additional parameter then gives models that are defined on a rather short slice of AdS space, raising questions as to whether it is still reasonable to use the AdS/CFT dictionary directly.  

Since the reasons for worrying about the consistency of the matching to the weakly coupled region are rather general,
it is not implausible that problems may also arise even for conserved
currents. Specifically, there are some subtleties associated with vector
currents that we will discuss in a forthcoming publication.

In holographic models that do not systematically deal with
scale-dependence, the natural way to evade the problems discussed in
this Letter is to give up on matching the holographic models to QCD in
the asymptotic regime, and simply fit all of the parameters of the 5D
models phenomenologically.  In the analysis above, this would correspond to fitting $a m_q$ and $\sigma/a$ to data, without computing $a$ separately, and giving up on the identification of $m_q$ as the source of the QCD operator $\bar{q}_R q_L$.  This has the disadvantage of losing many of
the theoretical connections to QCD.  This scenario is not ideal, but
it is not obviously inconsistent.  While it is conceivable that it may
be possible to treat scale-dependence consistently by modifying the
holographic models, it is clear that the issues that we have discussed
in this Letter must be addressed in bottom-up holographic models of
QCD. 

\begin{acknowledgements}
We gratefully acknowledge the support of the United States Department of Energy.  We thank J.~Erlich, N.~Evans, L.~Glozman, S.~Gubser and A.~Nellore for useful conversations, and M.~Stephanov and D.~Son for pointing out the subtlety associated with normalizing the fields. 
\end{acknowledgements}

\end{document}